\documentclass[a4paper]{jpconf}
\usepackage{graphicx}
\usepackage{aasjournals}
\usepackage{amsbsy}

\usepackage{color}

\newcommand{\Ni}[1]{\ensuremath{{}^{#1}{\rm Ni}}}
\newcommand{\Co}[1]{\ensuremath{{}^{#1}{\rm Co}}}
\newcommand{\Fe}[1]{\ensuremath{{}^{#1}{\rm Fe}}}
\newcommand{\code}[1]{\textsc{#1}}
\newcommand{\FLASH}{\code{flash}}
\newcommand{\PARAMESH}{\code{paramesh}}
\newcommand{\pv}{\ensuremath{\phi}}
\newcommand{\bvec}[1]{\ensuremath{\boldsymbol{#1}}} 
\newcommand{\grad}{\bvec{\nabla}} 

\newcommand{\At}{{\rm At}}
\newcommand{\ee}[1]{\ensuremath{\times 10^{#1}}}
\newcommand{\cdens}{\rho_{c}}

\newcommand{\unitspace}{\ensuremath{\,}}
\newcommand{\usp}{\unitspace}
\newcommand{\numberspace}{\ensuremath{\;}}
\newcommand{\nsp}{\numberspace}
\newcommand{\unitstyle}[1]{\ensuremath{\mathrm{#1}}}
\newcommand{\power}[2]{\ensuremath{{#1}^{#2}}}

\newcommand{\centi}{\unitstyle{c}}

\newcommand{\meter}{\unitstyle{m}}

\newcommand{\cm}{\centi\meter}
\newcommand{\gram}{\unitstyle{g}}

\newcommand{\grampercc}{\gram\usp\power{\cm}{-3}} 

\begin{document}
\title{On Simulating Type Ia Supernovae}

\author{A.~C. Calder$^{1,2}$, B.~K.~Krueger$^1$, A.~P.~Jackson$^{1,3}$, 
D.~M.~Townsley$^4$, E.~F.~Brown$^{5,6}$, and F.~X.~Timmes$^{7,6}$}

\address{$^1$ Department of Physics and Astronomy, Stony Brook University, Stony Brook, NY 11794-3800}
\address{$^2$ New York Center for Computational Sciences, Stony Brook University, Stony Brook, NY, 11794-3800}
\address{$^3$ present address: Laboratory for Computational Physics \& Fluid Dynamics,
Naval Research Laboratory, Washington, DC, 20375}
\address{$^4$ Department of Physics and Astronomy The University of Alabama, Tuscaloosa, AL, 35487-0324}
\address{$^5$ Department of Physics and Astronomy, Michigan State University, East Lansing, MI 48824-2320}
\address{$^6$ The Joint Institute for Nuclear Astrophysics, Notre Dame, IN 46556-5670}
\address{$^7$ School of Earth and Space Exploration, Arizona State University, Tempe, AZ, 85287-1404}

\ead{acalder@mail.astro.sunysb.edu}

\begin{abstract}
Type Ia supernovae are bright stellar explosions distinguished
by standardizable light curves that allow for their use as
distance indicators for cosmological studies. Despite their highly
successful use in this capacity, the progenitors of these events
are incompletely understood. We describe simulating type Ia
supernovae in the paradigm of a thermonuclear runaway occurring
in a massive white dwarf star. We describe the multi-scale
physical processes that realistic models must incorporate and the
numerical models for these that we employ. In particular, we describe
a flame-capturing scheme that addresses the problem of turbulent
thermonuclear combustion on unresolved scales. We present the
results of our study of the systematics of type Ia supernovae
including trends in brightness following from properties of the
host galaxy that agree with observations. We also present
performance results from simulations on leadership-class
architectures.
\end{abstract}

\section{Introduction}

\subsection{Type Ia Supernovae}

Type I supernovae are bright stellar explosions characterized by
a lack of hydrogen in the observed spectrum.  The type Ia sub-classification
depends on the observation of a specific silicon 
line~\cite{filippenko:optical,hillebrandt.niemeyer:type}.
These events are thought to be the result of a thermonuclear
explosion consuming roughly one and a half solar masses of
degenerate stellar material composed principally of C and O.
The peak brightness of a type Ia supernova is set not by the 
explosion energy, but 
by the synthesis in the explosion of radioactive \Ni{56}, which decays 
via the chain \Ni{56} to \Co{56} to \Fe{56}
releasing the observed energy.
The vast majority of
type Ia supernovae obey a correlation in which the peak brightness
is positively correlated with the 
timescale over which the lightcurve decays from its maximum.  This
``brighter is broader'' relation is known as the Phillips 
relation~\cite{phillips:absolute} and it allows the 
peak brightnesses to be calibrated so that these events may be treated 
as ``standard candles" for determining distances.  The correlation is 
understood physically as stemming from having both the luminosity and 
opacity being set by the mass of \Ni{56} synthesized in the 
explosion~\cite{arnett:type,pinto.eastman:physics,Kasen2007On-the-Origin-o}. 
This relation has been exploited to make type Ia supernovae the premier 
distance indicators for cosmological studies.

While it is accepted that type Ia supernovae result from a thermonuclear 
runaway as described above, the progenitor systems of these events remain
the subject of debate. Motivated largely by interest in 
cosmology, observational campaigns are gathering information about
these events at an unprecedented rate. In particular, observations
are finding a surprising range in the variation of the intrinsic brightness 
of these events, including very bright events that suggest more 
material burned than could originate from a single white 
dwarf~\cite{howell+06,scalzo+10,yuan+10,tanaka+10}. Contemporary 
research explores the efficacy of several progenitor systems
for producing events with a range of brightnesses. 
The Phillips relation explains the first-order variations in 
peak brightness, but current research also aims to understand 
higher-order effects on the variation of the light curves
and the physics behind the Phillips relation.

The progenitor system we assume is that of a single white dwarf 
composed principally of carbon and oxygen that has gained mass from a
companion star~\cite{hillebrandt.niemeyer:type}.
The process of the white dwarf gaining mass from the companion
is known as accretion. As the white dwarf gains mass, it is 
compressed and the temperature
increases, which ignites thermonuclear reactions. Initially the
reactions proceed relatively slowly, which drives convection in the
core~\cite{woosleyetal+04,zingale+09}. As the temperature rises, 
the thermonuclear reactions proceed faster and faster, and 
eventually a subsonic flame is born that will eventually
disrupt the star. Models that incorporate a transition from
the subsonic deflagration to a supersonic detonation, allowing
some expansion of the star prior to the detonation,
best match some spectral features
and the abundance stratification observed in the 
ejecta~\cite{Khokhlov1991Delayed-detonat,HoefKhok96,kasen_ca}.
Our models assume this deflagration-to-detonation transition (DDT)
paradigm.

\subsection{Trends in Observations of Type Ia Supernovae}

Many contemporary observations address correlations between the event
and properties of the host galaxy such as its composition, age,
and mass. Of particular interest are
correlations between the brightness of an event and the isotopic
composition and age, measured by the intensity of
star formation, of the host galaxy.
The proportion of material that has previously been processed in
stars (i.e.\ elements other than hydrogen and helium, which are
collectively referred to as ``metals'') is a measurable property
of the galaxy.  The  relative abundance of these elements is referred
to as the galaxy's ``metallicity.'' The presence of these elements in 
a progenitor white dwarf influences the outcome of the explosion by 
changing the path of nuclear burning, which influences the amount of 
$^{56}$Ni synthesized in an event~\cite{timmes.brown.ea:variations}. 
Because the decay of $^{56}$Ni powers the light curve, metallicity 
can therefore directly influence the brightness of an event. 
Metallicity can also influence the outcome of an explosion in other ways, 
including changes in the structure of the white dwarf that result from 
metallicities influence on stellar evolution, sedimentation within the 
white dwarf, the nuclear flame speed, and additional sources of 
opacity~\cite{DomiHoefStra01}.

Observational results to date are consistent 
with a shallow dependence of brightness on metallicity, with dimmer 
events in metal-rich galaxies, but are unable demonstrate a conclusive 
trend~\cite{GallGarnetal05,gallagheretal+08,neilletal+09,howelletal+09}.
Determining the metallicity dependence is challenging because the
effect appears to be small, it is difficult to measure, and
there are systematic effects associated with the
mass-metallicity relationship within galaxies~\cite{gallagheretal+08}.
This effect is also difficult to decouple from the apparently
stronger effect of the age of the parent stellar population on the mean
brightness of type Ia supernovae~\cite{gallagheretal+08,howelletal+09,Krueger2010On-Variations-o}.
Howell, et al.~\cite{howelletal+09} note that the scatter in brightness of this observed
relation is unlikely to be explained by the effect of metallicity.

The composition of a galaxy changes as stars form, consume hydrogen,
and produce metals. Newly-formed galaxies are rich in hydrogen gas and 
undergo a period of intense star formation. Observations target correlations 
between the brightness of an event and the age of a galaxy measured
by the delay time (the elapsed
time from the period of intense star formation). Some results indicate
that the dependence of the Ia rate on delay time
is best fit by a bimodal distribution
with a prompt component less than 1~Gyr after star formation and a tardy
component several Gyr later~\cite{MannucciEtAl06,RaskinEtAl09}.
Other studies only indicate a correlation between the delay time and the
brightness of type Ia supernovae, with dimmer events occurring at longer delay
times~\cite{gallagheretal+08,howelletal+09,neilletal+09,BrandtEtAl10}.

\section{Methodology}

Our investigation into type Ia supernovae proceeds by isolating
facets of the problem and subjecting these to study via statistically
well-controlled ensembles of multidimensional simulations.  We are trying 
to understand the physical mechanism of the explosion and isolating one 
effect at a time allows us to understand these processes. The parts of the 
problem we consider are properties of the progenitor white dwarf (such as 
structure and composition) that follow from properties the host galaxy 
(such as age and metallicity). The goal is to identify and quantify
systematic trends in the brightness of events that follow from properties of 
the host galaxy. There are many possible systematic effects (outlined 
in \cite{townetal09}), and for each study we attempted to isolate all 
but one effect.  Eventually, we hope to consider the interdependence of 
all of these effects in the construction of the full theoretical picture.
Another feature of the type Ia supernova problem is that the explosion
depends sensitively on the initial conditions, which are largely unknown.

We construct parametrized, hydrostatic massive white dwarf progenitor models 
with a variety of thermal and compositional structures thought to follow from 
the properties of the host galaxy and the accretion history~\cite{jacketal}. 
The simulations are performed with a customized version of the \FLASH\ simulation 
code~\cite{Fryxetal00,calder.curtis.ea:high-performance,calder.fryxell.ea:on} that 
includes a model flame and energetics scheme. We perform suites of simulations in 
which we vary properties of the progenitor white dwarf. The suites of simulations 
are controlled by a theoretical framework we developed that allows for statically 
well-controlled studies to determine any systematic effects due to properties of the 
progenitor that may follow from properties of the host galaxy.
 
Simulations with \FLASH\ provide the bulk energetics of the 
explosion and an estimate of the yield of an event. 
We calculate detailed nucleosynthetic yields from the 
simulations by post-processing the thermodynamic 
histories of Lagrangian tracer particles embedded in the flow
with a detailed nuclear network~\cite{brown.calder.ea:type,townetal10}. 

In the subsections below, we highlight the flame model, sub-grid-scale
turbulence and turbulence-flame interaction models, and the statistical 
framework we employ. Complete details of the methodology can be found 
in previously published 
results~\cite{brown.calder.ea:type,Caldetal07,townsley.calder.ea:flame,townetal09,townetal10}.

\subsection{Model Flame and Energetics Scheme}

The tremendous range in length between the white dwarf
radius ($\sim 10^9\nsp\cm$) and the laminar nuclear flame width ($\approx
[\kappa T/(\rho \varepsilon)]^{1/2} < 1\nsp\cm$ for typical values 
of the thermal conductivity $\kappa$, density $\rho$,
temperature $T$, and mass-specific heating rate $\varepsilon$), 
prohibits direct numerical simulation of type Ia supernovae.
The problem is so severe that even an adaptive
mesh simulation cannot resolve the actual diffusive flame front.
For these reasons,  simulations must rely on 
some sort of ``model'' flame. 
The structure of a one-dimensional laminar burning front is reasonably well
understood~\cite{ChamBrowTimm07}, 
but the interaction of turbulence with this very
thin flame front introduces small-scale structure in a way that is
not well understood, even for flames in the laboratory. Capturing the
net effect of this unresolved flame structure is necessary for a realistic
model.

Thermonuclear flames in \FLASH\ are tracked with an advection-diffusion-reaction 
scheme~\cite{Khok95,VladWeirRyzh06,townsley.calder.ea:flame} that propagates
an artificially broadened model flame.
This model is based on the
evolution of a reaction progress variable $\phi$, where $\phi=0$ indicates
unburned fuel and $\phi=1$ indicates burned ash, satisfying the
advection-diffusion-reaction equation
\begin{equation}
  \label{eq:ard}
  \partial_t \pv + \bvec{u}\cdot\grad \pv = \kappa \nabla^2 \pv + 
\frac{1}{\tau} R\left(\phi\right).
\end{equation}
Here $R(\phi)$ is a single chosen function, used everywhere,
and $\kappa$ and $\tau$ are parameters that are tuned locally so that the 
reaction front propagates at the physical speed of the real flame in 
that region~\cite{ChamBrowTimm07} and is just wide
enough to be resolved in our simulation.  We use a modified version of the
KPP reaction rate discussed by~\cite{VladWeirRyzh06}, in which
$R\left(\phi\right) \propto \left(\phi-\epsilon\right)\left(1-\phi+\epsilon\right)$, where 
$\epsilon \simeq 10^{-3}$, which is acoustically quiet and gives a unique flame 
speed~\cite{townsley.calder.ea:flame}.

In addition to the model flame, a realistic supernova model
must accurately describe  the energy release at the front
and in the burned material, which is in a dynamic equilibrium.
We performed a detailed study of the nuclear processes occurring in
a flame in the interior of a white dwarf and developed an efficient and 
accurate scheme for numerical simulation~\cite{Caldetal07,townsley.calder.ea:flame}. Tracking
many nuclear species is computationally prohibitive, so 
simulations reproduce the energy release with an abstracted model with a 
tabulated state of the burned material produced from earlier
directed simulations.
This method accurately captures the thermal history of the material 
as it burns and evolves, which enables embedded particles to obtain 
accurate Lagrangian density and temperature histories. Detailed
abundances can then be recovered by post-processing these time 
histories with a nuclear network including hundreds of nuclides.
The nuclear processing can be well approximated as a three-stage process:
Initially carbon is consumed, followed by oxygen, which creates a 
mixture of Si group and light elements that is in 
quasi-statistical equilibrium~\cite{bodclafow1968,woosleyarnettclayton73} (also 
known as nuclear statistical quasi-equilibrium~\cite{ifk1981,khok1981,khok1983}); finally 
the Si group nuclei are converted to 
Fe group, reaching full NSE. In both of these equilibrium states,
the capture and creation of light elements (via photodisintegration) is
balanced, so that energy release can continue by changing the relative
abundance of light (low nuclear binding energy) and heavy (high nuclear
binding energy) nuclides, an action that releases energy as burned 
material rises and expands. Neutronization in the flame and
dynamic ash, mainly via electron captures, is explicitly accounted for with 
tabulated rates from NSE calculations with 443 nuclides that include
modern weak reaction rates~\cite{langanke.martinez-pinedo:weak}. Complete
details of the NSE calculations may be found in~\cite{SeitTownetal09} and
the details of the implementation in our simulations may be found in
\cite{Caldetal07,townsley.calder.ea:flame}.

\subsection{Sub-grid-scale Models}

The model flame with its advection-diffusion-reaction scheme requires
an input flame speed. For the one-dimensional laminar case, flame speeds 
are readily available from direct numerical 
simulations~\cite{timmes_1992_aa,ChamBrowTimm07}
In the multidimensional case, turbulence on 
unresolved scales boosts the flame speed. Realistic simulations 
must therefore include a method for determining the effect 
of this turbulence-flame interaction (TFI).

Initially we employed a method proposed by Khokhlov~\cite{Khok95},
that assumes perturbations to the flame front are dominated by the
Rayleigh-Taylor instability.  The method 
sets the flame front speed to $s=\max(s_\ell, 0.5 \sqrt{\At\,gm\Delta})$
where $s_\ell$ is the laminar flame speed, $\At=(\rho_{\rm fuel}-\rho_{\rm
ash})/(\rho_{\rm fuel}+\rho_{\rm ash})$ is the Atwood number, $g$ is the
local gravity, $\Delta$ is the grid resolution, and $m$ is a calibrated
constant.  This prescription effectively keeps the flame speed above some lower
limit that depends on the resolution of the simulation.~\cite{townsley.calder.ea:flame}.  
Another approach that accounts for both unresolved Rayleigh-Taylor instability
and background turbulence, originally proposed by Niemeyer and 
Hillebrandt~\cite{NiemHill95} and developed in detail
by Schmidt et al.~\cite{Schmetal06a,Schmetal06b}, relies on 
a dynamic measure of the local turbulent energy on sub-grid scales to enhance 
the local flame front propagation. The model sets the flame speed 
to $s = \sqrt{s_\ell^2 + C_tq^2}$, where $q$ is a velocity that characterizes 
the sub-grid turbulence energy content and $C_t$ is a constant taken to be 4/3~\cite{Poch94}.

\begin{figure}
\centering
\includegraphics[width=5.5in,angle=0]{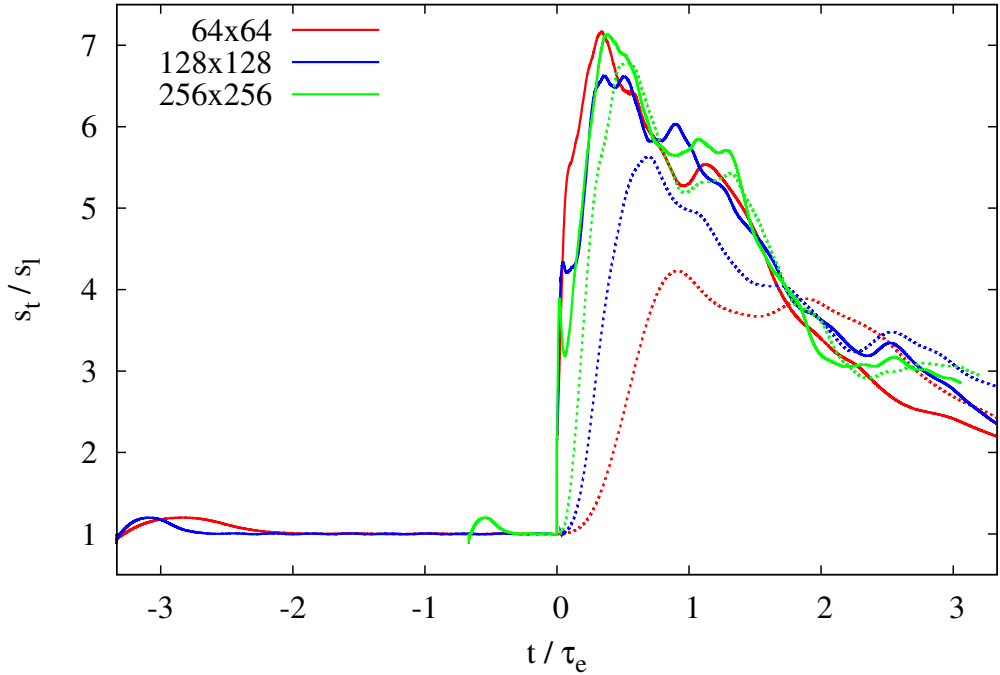}
  \caption{
    \label{fig:tfi}
   Convergence properties with resolution, with and without a TFI model.
   Shown is the ratio of the turbulent flame speed to the laminar flame
   speed as a function of the simulation time scaled by the turbulence eddy turnover 
   time at $t = 0$, $\tau_e$, in a channel with decaying turbulence.  Three different 
   resolutions are shown using a TFI model based on Charlette et al.~\cite{Charetal02a}
   (solid lines) and without a TFI model (dotted lines). Convergence with resolution is 
   achieved with the TFI model.
  }
\end{figure}

We have recently implemented a TFI model based on Colin et al.~\cite{Colietal00} 
and Charlette et al.~\cite{Charetal02a} that 
reasonably predicts the behavior of turbulent
flames in terrestrial experiments~\cite{SarlBeneRuss10}. The method
utilizes a local, instantaneous measure of the turbulence, in contrast
to the dynamic turbulence model employed by Schmidt et al.~\cite{Schmetal06a}. In addition,
the relation between the turbulent flame speed and the turbulent intensity
is more carefully considered. Particular care has been taken to ensure
that the turbulent flame speed predicted by the TFI model shows good
convergence properties with resolution for the case of Kolmogorov turbulence. 
Figure~\ref{fig:tfi} shows the
ratio of the turbulent flame speed to the laminar flame speed 
for a flame propagating in a channel with decaying turbulence having an initial
eddy turnover timescale
$\tau_e$. The ratio of the turbulent flame speed to the laminar flame speed 
is plotted as a function of simulation time scaled by $\tau_e$.
The solid lines show results from the TFI model following~\cite{Charetal02a}, 
while the dotted lines show results without the inclusion of a TFI 
model. Three tests were performed with cubic cells that cover the 
channel cross section with 64x64 cells (red), 128x128 cells (blue), 
and 256x256 cells (green) to demonstrate the necessity of a TFI 
model to obtain convergence with resolution.

\subsection{Statistical Framework}

The statistical framework we developed is designed to explore systematic 
trends in the brightness of type Ia supernovae.  The ideas, however, are 
general and the same approach may be taken for other multi-physics 
applications. Our approach allows the evaluation of the average dependence 
of properties of the model on underlying parameters by constructing a 
theoretical sample based on probabilistic initial conditions.  

For our type Ia supernova research, we utilize two- and three-dimensional 
simulations in the DDT paradigm within the framework
to evaluate the average dependence of the brightness of an event
on parameters such as composition of the progenitor. The
theoretical sample is based on probabilistic initial ignition conditions
for the deflagration. Such sample-averaged dependencies are particularly 
important to the type Ia supernova problem because we seek to understand 
how models may explain features of the observed sample, particularly samples 
generated by large dark energy surveys utilizing type Ia supernovae as distance 
indicators.

The theoretical sample is constructed to represent statistical properties of
the observed sample of type Ia supernovae such as the mean inferred $\Ni{56}$
yield and its variance. Within the DDT paradigm, 
the variance in $\Ni{56}$ yields can be explained by the development of fluid 
instabilities during the deflagration phase of the explosion. Choice of 
the initial configuration of the flame  influences the growth of fluid
instabilities, which  result in varying amounts of $\Ni{56}$ synthesized
during the explosion. While there are many sources of uncertainty, particularly 
in the progenitor white dwarf's composition and structure, variations of the 
initial configuration of the flame can be used to introduce fluid instabilities 
that produce the variance of $\Ni{56}$ yields seen in observations.

The physical initial conditions at ignition are not well known and most 
likely involve scales our simulations cannot resolve. Our initial conditions 
consist of an initially burned region that is resolved on the simulation grid.
We found that perturbing a spherical flame surface (with radius $r_0 = 150$ km) 
with spherical harmonic modes ($Y_l^m$)
between $12 \leq l \leq 16$ with random amplitudes ($A$) normally
distributed between $0-15$ km and, for three-dimensions, random phases ($\delta$)
uniformly distributed between $\mbox{-}\pi$ and $\pi$ best characterized the
mean inferred $\Ni{56}$ yield and sample variance from 
observations~\cite{townetal09}. We write the perturbation as
\begin{eqnarray}
\label{eq:init_surf}
r(\theta) &=& r_0 + \sum_{l=l_{\rm min}}^{l_{\rm max}} A_l Y_l(\theta) \\
r(\theta,\phi) &=& r_0 + \sum_{l=l_{\rm min}}^{l_{\rm max}} \sum_{m=-l}^l 
\frac{A_l^m e^{i\delta_l^m}}{\sqrt{2l+1}} Y_l^m(\theta,\phi)
{\rm .}
\end{eqnarray}
With a suitable random-number generator, a
sample population of progenitor WDs is constructed by defining the initial
flame surface for a particular progenitor.

Our theoretical framework takes advantage of the fact that the outcome of
an explosion is very sensitive to the initial conditions to produce the
sample population. We note that other aspects of the problem such as the
DDT density can significantly influence the $\Ni{56}$ yield as well. The 
mechanism by which a DDT occurs is poorly understood, but it might also
contribute to the variation seen in the observations.

\section{Performance Results}

\FLASH\ is a parallel, adaptive-mesh simulation code for 
multidimensional compressible reactive flows in astrophysical 
environments.  At its heart is an explicit hydrodynamics method, 
and \FLASH\ also includes solvers for the Poisson equation of self-gravity 
and the advection-diffusion-reaction scheme for propagating a model
flame as described above. \FLASH\ also provides an equation of state 
for a degenerate ionized plasma to describe stellar material.
\FLASH\ uses a customized version of the \PARAMESH\ 
library~\cite{macneice.olson.ea:paramesh,macneice.olson.ea:paramesh*1} to 
manage a block-structured adaptive grid, adding resolution elements in areas
of complex flow. \FLASH\ and \PARAMESH\ use MPI for parallel communication.

\begin{figure}
\centering
\includegraphics[width=5.0in,angle=0]{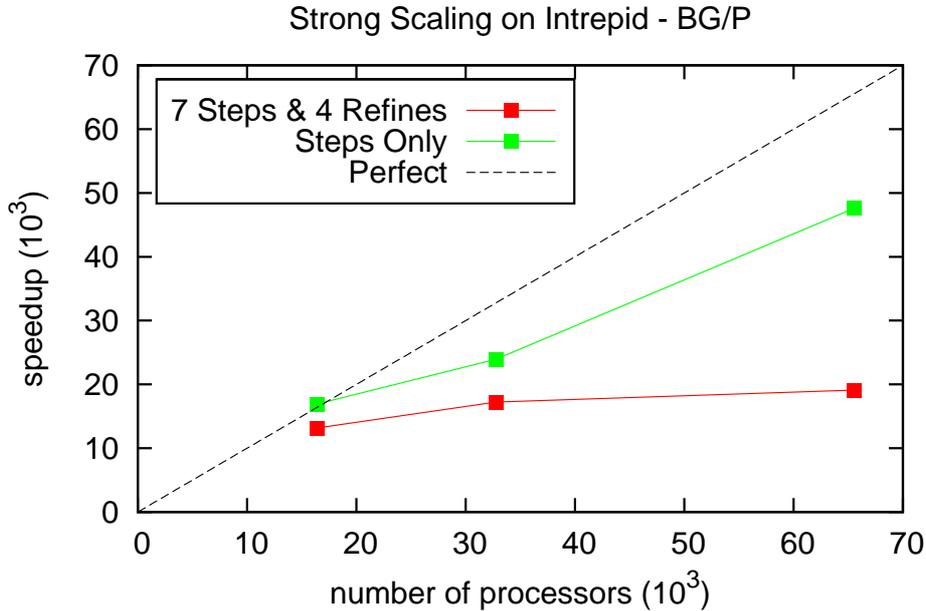}
  \caption{
    \label{fig:speedup}
   Strong scaling results for the ANL BG/P machine illustrating parallel speedup. 
   Plotted is the speedup vs.\ number of processors for 16384, 32768, and 65536 
   processors. Also shown is the corresponding ideal scaling case.
  }
\end{figure}

\FLASH\ has always demonstrated almost perfect weak scaling, winning the SC2000
Gordon Bell Prize (special category)~\cite{calder.curtis.ea:high-performance},
for scaling to 6420 processors of the Intel ASCI Red machine at LANL in 2000.
As an example of strong scaling for production simulations, 
we present a performance test run on the BG/P machine at ANL. The
results shown in Figure~\ref{fig:speedup} are for a three-dimensional simulation under
conditions typical for a production research run. 

Simulations of type Ia supernovae utilize adaptive mesh refinement
to track critical features of the simulation, particularly the location
of the flame, at the highest resolutions. 
As the star explodes, the
size of the simulation increases with time as the
burned volume of the star increases. Accordingly, 
the proportion of simulation domain at the highest resolution grows, drastically
increasing the size of the simulation. For this reason, the timing for the 
scaling study considered only a small part of the actual evolution.
The test measured the amount of wall clock time 
taken to evolve a supernova simulation for a fixed period of simulation
time (corresponding to 7 time steps, with mesh adaptations and
load balancing every two time steps for a total of 4 adaptations).  
The study was run on 16384, 32768, and 65536 processors,
and the simulation was for a
late-stage of the deflagration, and is thus representative of a fully
developed simulation. Figure~\ref{fig:speedup} illustrates the parallel speedup,
\begin{displaymath}
{\mathrm{speedup}} = \frac{16384 t_{16384}}{t_N}
\end{displaymath}
where $t_N$ is the simulation time run on $N$ = 16384--65536 processors.
In this case, perfect scaling corresponds to the number of processors.
Work is needed to improve the cost of mesh refinement.

\section{Astrophysical Results}

In Townsley et al.~\cite{townetal09} we investigated the effect of 
the metallicity of the host galaxy by assuming it determined the 
initial neutron excess of the progenitor white dwarf.  Weak 
interactions during nuclear burning, particularly electron capture, 
lead to the production of neutron-rich elements. The neutron excess 
of these elements is thought to drive the explosion yield toward 
stable iron-group elements. As a result, there is relatively less
radioactive $^{56}$Ni in the yield of iron-group elements, which 
results in a dimmer event~\cite{timmes.brown.ea:variations}. 
Thus the introduction of neutron-rich metals into the progenitor white 
dwarf is thought to influence the brightness of an event by influencing 
the $^{56}$Ni yield.

We investigated the role of metallicity by introducing $^{22}$Ne into the
progenitor white dwarf as a proxy for neutron-rich metals.
The presence of $^{22}$Ne influences the progenitor structure, the energy
release of the burn, and the flame speed.  The study was designed to measure
how the $^{22}$Ne content influences the competition between rising plumes
and the expansion of the star, which determines the yield.
We performed a suite of 20 two-dimensional DDT simulations varying only the
initial $^{22}$Ne in a progenitor model, and found a
negligible effect on the pre-detonation expansion of the star
and thus the yield of iron-group elements.  The neutron excess sets the
amount of material in NSE that favors stable iron-group elements over
radioactive $^{56}$Ni. Our results were consistent with earlier
work calculating the direct modification of $^{56}$Ni mass from
initial neutron excess~\cite{timmes.brown.ea:variations}.

We expanded the study of Townsley et al.~\cite{townetal09} to include the 
indirect effect of metallicity in the form of the $^{22}$Ne mass fraction 
through its influence on the density at which the DDT takes place in Jackson 
et al.~\cite{jacketal}.  The study consisted of 30 ``realizations'' or sets 
of initial conditions (the randomized sample), and for each performed
two-dimensional simulations with 5 transition densities 
between $1-3\times10^7$~g~cm$^{-3}$ for a
total of 150 simulations.  We found a quadratic dependence of the iron-group
yield on the log of the transition density, which is determined by the
competition between rising unstable plumes and stellar expansion. By then 
considering the effect of metallicity on the transition density, we found 
that the iron-group yield decreases slightly with metallicity, but that the 
ratio of the $^{56}$Ni yield to the overall iron-group yield does not change
significantly.  Observations testing the dependence of the yield on metallicity
remain somewhat ambiguous, but the dependence we found is comparable to that
inferred from~\cite{bravo10}. We also found that the scatter in the results
increases with decreasing transition density, and we attribute this increase in
scatter to the nonlinear behavior of the unstable rising plumes.

In Krueger et al.~\cite{Krueger2010On-Variations-o} we investigated
the effect of central density on the explosion yield.
We performed a suite of simulations from 30 realizations, and for
each performed two-dimensional simulations with 5 central densities
between  $1-5\times10^9$~g~cm$^{-3}$.
We found that the overall production of iron-group material did not
change, but there was a definite trend of decreasing $^{56}$Ni
production with increasing progenitor central density (consistent
with earlier studies~\cite{Iwametal99,Bracetal00}).
We attribute this result to higher rates of weak interactions
(electron captures) that produce a higher proportion
of neutronized material when the burning occurred at higher density.  
Similarly to the influence of metals, more neutronization means less 
symmetric nuclei like $^{56}$Ni, and, accordingly, a dimmer event. 

This result may explain the observed decrease in the brightness 
of events with increasing age measured as delay time from star formation. 
For the accreting white dwarf progenitor, only a narrow window in 
the range of possible accretion rates will produce a massive 
progenitor in which carbon can be centrally ignited, 
avoiding far off-center ignitions and subsequent gravitational
collapse due to high electron-capture rates. Spherically symmetric
models generally find central densities of
$1.8 - 13.0 \ee{9}\nsp\grampercc$, depending on
the mass of the accreting WD~\cite{nomoto_1985_aa,hernanz_1988_aa,bravo_1996_aa}.
This narrow window of rates implies a similar accretion duration
for all progenitor systems. 

\begin{figure}[htb]
\centering
\includegraphics[width=5.0in]{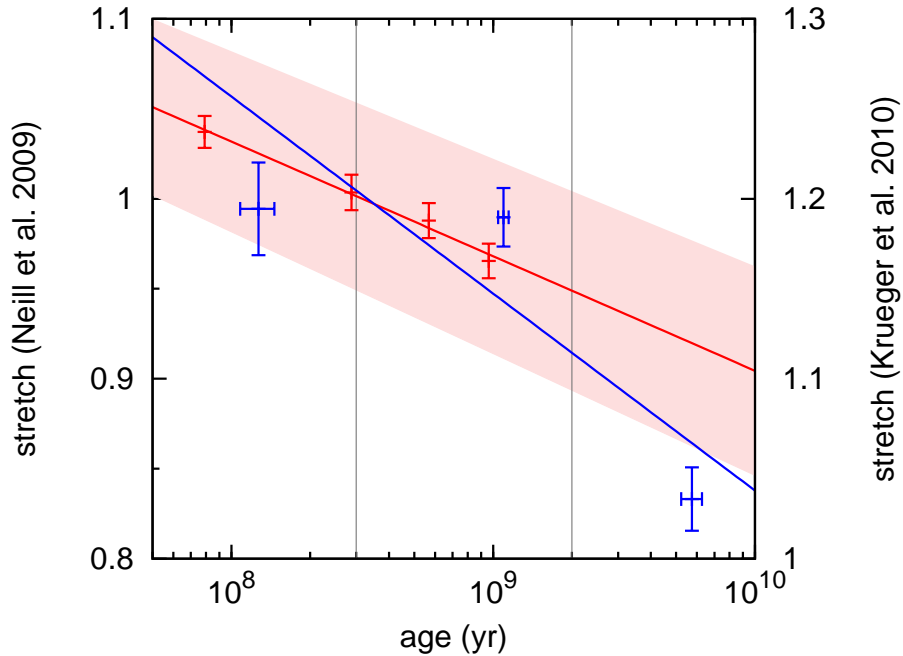}\hfill%
\caption{Plots of stretch ($s$)~\cite{howelletal+09}, a measure related to 
brightness~\cite{Goldhaber2001Timescale-Stret}, vs.\ age, comparing our trend
in brightness with progenitor age to observational results.
In red are results from our study, based on variations in $\cdens$, along with
the standard deviation (shaded region). In blue are the binned and averaged
points from Figure 5 of~\cite{neilletal+09}, along with a best-fit line.
Note the different scales: the overall offset to larger stretch in the
simulations is due to the choice of DDT density.
Adapted from~\cite{Krueger2010On-Variations-o}.
\label{f.stretch}
}
\end{figure}

Given that all progenitor systems seem to have the same duration
of accretion, differences in progenitor age must then follow from 
differences in the time scale of evolution prior to the  onset of accretion.
If there is a long period of cooling before the onset of mass transfer,
the central density of the progenitor will be higher when the core
reaches the carbon ignition temperature~\cite{Lesaffre2006The-C-flash-and},
thereby producing less $^{56}$Ni and thus a dimmer event.  
Using the results of Lesaffre~\cite{Lesaffre2006The-C-flash-and}, we were able to relate our
results to the progenitor age and compare our results to
observations, shown in Figure~\ref{f.stretch}.
In addition, in this study we found
considerable variation in the trends from some realizations, stressing the
importance of statistical studies~\cite{scidac}.

\section{Summary and Conclusions}

We presented a snapshot of our research into the systematics of
type Ia supernovae at the time of the Conference on Computational
Physics held in Gatlinburg, TN, in the fall of 2011. We outlined our 
methodology and presented some highlights from our research into the
systematics of type Ia supernovae. Our goal for this research it to
develop sophisticated models that can reliably address issues like
the intrinsic scatter of type Ia supernovae, an issue critical for
the use of these events as cosmological distance indicators. Our
contemporary research focuses on identifying trends in the brightness
properties of the progenitor white dwarf that follow from the properties
such as age and composition of the host galaxy. We developed a
theoretical framework allowing study of the systematics of type
Ia supernovae via statistically well-controlled suites of 
simulations~\cite{townsley.calder.ea:flame}, and we applied this
method to studies of the composition and structure of the progenitor
white dwarf.

Our results suggest that the direct effect of metallicity on the outcome of our 
multidimensional explosion models is very much in keeping with previous 
one-dimensional studies, e.g.~\cite{timmes.brown.ea:variations}. A suite of 
simulations showed a negligible effect on the expansion of the white dwarf
prior to the detonation, and hence the yield~\cite{townetal09}.  We find that 
considering the indirect effect of metallicity on flame speeds and the deflagration 
to detonation transition produces a stronger trend. We found a quadratic dependence 
of the yield on the log of the transition density, which is determined by the 
competition between plume rise and stellar expansion during the deflagration phase. 
By considering the effect of metallicity on the transition density, we obtained a 
relationship between brightness and metallicity~\cite{jacketal}.

Our results from variations of the central density of the  progenitor provide
a theoretical explanation of the observed trend that type Ia supernovae from older 
host galaxies are systematically dimmer. Previous
work~\cite{Bravo1990, RoepGiesetal06, fisheretal10,
Hoeflich2010SecondaryParametersSNeIa, seitenzahletal11}
addressed the question of the impact of central density on
brightness, with different studies reaching different conclusions.
We found a strong trend of decreasing brightness with increasing 
central density due to increased rates of weak interactions that
drive the burning toward more neutron rich products. By relating these results to
the accretion history of the progenitor, we were able to obtain
a relationship between brightness and age of the system~\cite{Krueger2010On-Variations-o}.  

A result drawn from all of the studies is dependence of the problem on the
morphology of the deflagration. The deflagration phase of type Ia supernovae is 
strongly influenced by fluid instabilities and has a very nonlinear evolution.
These results stress the necessity of a statistical study in order to capture
the true trend. 

Our conclusion is that we are beginning to understand the systematics of 
these events, a necessary step prior to addressing the issue of the intrinsic
scatter. Future work will continue in this direction, with a study of the
effect of the C/O ratio in the progenitor in progress. We also continue to
refine and develop our models, particularly the sub-grid-scale turbulence
and turbulence-flame interaction models. We have explored much of the parameter
space of white dwarf progenitors and will perform targeted three-dimensional
simulations to support our present results.

\ack

This work was supported by the Department of Energy through grants
DE-FG02-07ER41516, DE-FG02-08ER41570, and DE-FG02-08ER41565, and by NASA
through grant NNX09AD19G.  ACC also acknowledges support from the
Department of Energy under grant DE-FG02-87ER40317. DMT received support
from the Bart J. Bok fellowship at the University of Arizona for part of
this work. The authors acknowledge the hospitality of the Kavli Institute
PHY05-51164, during the programs ``Accretion and Explosion: the Astrophysics
of Degenerate Stars'' and ``Stellar Death and Supernovae.''  The software
used in this work was in part developed by the DOE-supported ASC/Alliances
Center for Astrophysical Thermonuclear Flashes at the University of Chicago.
This work was also supported in part by the U.S.\ Department of Energy,
Office of Nuclear Physics, under contract DE-AC02-06CH11357 and
utilized resources at the New York Center for Computational Sciences
at Stony Brook University/Brookhaven National Laboratory which is
supported by the U.S.\ Department of Energy under Contract No.
DE-AC02-98CH10886 and by the State of New York. 
Finally, the authors acknowledge and thank the organizers and committees 
of the Conference on Computational Physics held October 30 -- November 3, 2011 
in Gatlinburg, TN for putting together a delightful and informative
conference.

\section*{References}

\end{document}